\documentclass[12pt,epsf]{article}

\usepackage{graphicx}
\usepackage{epsfig}
\usepackage{amsmath}

\usepackage{epstopdf}
\usepackage{color}
\usepackage{amssymb}
\usepackage{verbatim}

\newcommand{\pd}[2]{\frac{\partial #1}{\partial #2}}

\newcommand{\dd}{\mathrm{d}}

\newcommand{\He}{\mathrm{H}}

\newcommand{\phat}{\hat{p}}

\newcommand{\that}{\hat{t}}
\newcommand{\zhat}{\hat{z}}

\newcommand{\Qhat}{\hat{Q}}
\newcommand{\Rhat}{\hat{R}}
\newcommand{\What}{\hat{W}}

\newcommand{\ttil}{\tilde{t}}

\newcommand{\Stil}{\tilde{S}}

\newcommand{\deltatil}{\tilde{\delta}}
\newcommand{\etatil}{\tilde{\eta}}

\newcommand{\od}[2]{\frac{\dd #1}{\dd #2}}

\title{A Mathematical Model of the Rainwater Flows in a Green Roof}

\author{
}\date{}

\begin{document}

\maketitle

\begin{center}
Catherine C Adley \\
Department of Chemical and Environmental Sciences, \\
University of Limerick, \\
Limerick, Ireland, \\

\

Mark J Cooker \\
School of Mathematics, University of East Anglia, \\
Norwich, NR4 7TJ, U.K., \\

\

Gemma L Fay \\
OCIAM, Mathematical Institute, \\
University of Oxford, \\
24-29 St. Giles', Oxford, OX1 3LB, U.K., \\

\

Ian Hewitt \\
OCIAM, Mathematical Institute, \\
University of Oxford, \\
24-29 St. Giles', Oxford, OX1 3LB, U.K., \\

\

Andrew A Lacey \\
Maxwell Institute for Mathematical Sciences, \\
and  School of Mathematical and Computer Sciences, \\
Heriot-Watt University, Riccarton, Edinburgh, EH14 4AS, U.K., \\

\

Niklas Mellgren \\
KTH, Department of Mechanics, \\
SE-100 44 Stockholm, Sweden, \\

\

Marguerite Robinson \\
Institut Catal\`{a} de Ci\`{e}ncies del Clima (IC3), C/Doctor Trueta
203, 08005 Barcelona, Catalunya, Spain, \\

\

and Michael Vynnycky \\
Mathematical Applications Consortium
for Science and Industry (MACSI), \\
Department of Mathematics \& Statistics,
College of Science \& Engineering, \\
University of Limerick, Limerick, Ireland.
\end{center}

\begin{abstract}
A model is presented for the gravity-driven flow of rainwater
descending through the soil layer of a green roof, treated as a
porous medium on a flat permeable surface representing an
efficient drainage layer. A fully saturated zone is shown to
occur. It is typically a thin layer, relative to the total soil
thickness, and lies at the bottom of the soil layer. This provides
a bottom boundary condition for the partially saturated upper
zone. It is shown that after the onset of rainfall, well-defined
fronts of water can descend through the soil layer. Also the
rainwater flow is relatively quick compared with the moisture
uptake by the roots of the plants in the roof. In separate models
the exchanges of water are described between the (smaller-scale)
porous granules of soil, the roots and the rainwater in the
inter-granule pores.
\end{abstract}

\section{Introduction}
Green roofs are becoming increasingly popular around the world.
The many benefits of a green roof include assistance in the
management of storm water, pollution control, building insulation and
recycling of carbon dioxide, in addition to being aesthetically
pleasing. A green roof is subject to various stresses from the
weather, in particular wind-loading, which we ignore in this
report, and rainfall: it is the flow, drainage and uptake of
rainwater that we model. An understanding of where the water goes
is essential to design a roof able to achieve sustained healthy
plants and loads that lie within the safe capacity of the
supporting structure.

The main focus of this paper is on the transport of water through
the structure of the green roof. Inadequate drainage can lead to the
undesirable occurrence of a fully saturated soil which will cut
off the air supply to the plants. Conversely, if the saturation
levels are too low plants will die from lack of water. Ideally a
degree of saturation that is less than eighty per cent should be
maintained at all times. Our goal is to model the distribution of
the degree of saturation through the depth of the soil layer, and
to see how it changes due to spells of rain, and under the
influence of moisture-uptake by plant roots.

This study was motivated by a problem brought to the 70th
European Study Group with Industry, held in Limerick in 2009.
The moisture input into the roof used later is, therefore,
based on Irish weather data.

\

\begin{figure}
\centering
\resizebox{13cm}{6.5cm}{\includegraphics{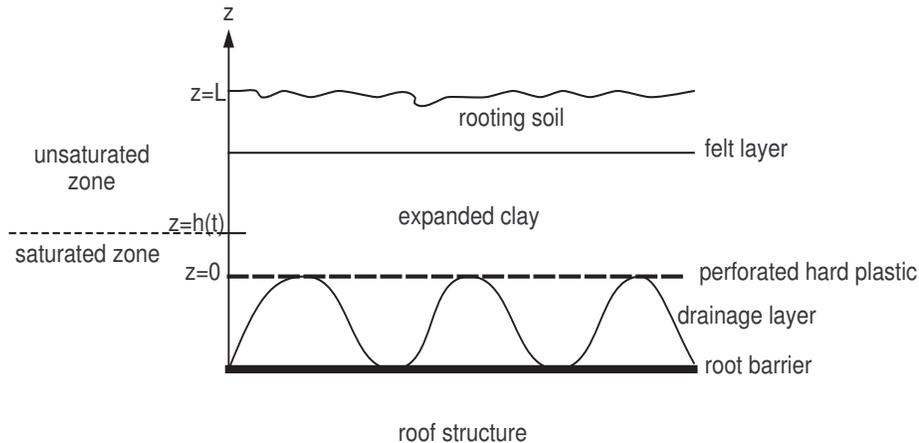}}\\
\caption{Green roof structure. Gravity is in the negative
$z$ direction.} \label{roof}
\end{figure}

The basic structure of a common green roof is shown in Fig.~\ref{roof}.
A waterproof root barrier protects the underlying roof
structure. A drainage layer sits atop this barrier. The typical
thickness of this layer is 8/15/20~mm depending on the type of
roof. The drainage layer has not been modelled in this
study, and any possible build-up of water there has been
disregarded. Instead, any water entering this layer is assumed,
perhaps unrealistically, to leave
the system. The soil and drainage layers are separated by a thin sheet
of perforated hard plastic containing holes approximately 2~mm in
diameter and spaced roughly 2~cm apart. There are two layers of soil at
the top of the structure separated by a layer of felt. A thin
layer ($<$~2~cm) of refined rooting soil contains the plant life,
mainly sedum for thinner roofs and, for thicker ones, low growing
grasses such as common bent grass and/or other plants, such as
cowslip and ladies bedstraw. Beneath the rooting soil are pellets
of lightweight expanded clay. This layer is 5-10~cm thick. Grain
sizes are typically $<$~2 mm for rooting soil and 4-8~mm for
expanded clay pellets.


\section{The Model}
We model the dynamics of water flow through the soil layer. We
consider a single soil layer with thickness $L\approx 10^{-1}$ m
and we ignore the presence of the felt layer. We assume that the
soil-drainage-layer interface is located at $z=0$ and the soil
surface at $z=L$. We consider two possible scenarios: (\textit{i})
the entire region $0\leq z\leq L$ is unsaturated, so that the soil
saturation $S$ is everywhere less than 1 and (\textit{ii}) a saturated
region $0\leq z\leq h$ lies at the bottom of the soil layer.
Note that the model as presented here is one-dimensional, and
represents a horizontal roof, but can be easily extended to two
(or three) dimensions, and to account for sloping roofs.


\subsection{The Unsaturated Region}
We first assume the entire region $0\leq z\leq L$ is unsaturated
($S<1$). The basic model for this region follows that outlined in
\cite{genuchten80} and \cite{roose04a}. The equation
for water flow in the unsaturated soil comes from
having local water flux measured upwards (in the positive $z$ direction) 
\begin{equation}\label{fluxA}
q = -  \left(D_{0}D(S)\frac{\partial S}{\partial z} + K_{0}K(S) \right)
\end{equation}
and employing the balance law
\begin{equation}\label{balanceA}
\phi \frac{\partial S}{\partial t} + \frac{\partial q}{\partial z} =
- R ,
\end{equation} to give the one-dimensional
Richards' equation (see \cite{fulford02} and \cite{richards31})
\begin{equation}\label{richards}
\phi \frac{\partial S}{\partial t}=\frac{\partial}{\partial
z}\left(D_{0}D(S)\frac{\partial S}{\partial z}+K_{0}K(S)\right)-R,
\end{equation}
where $S=S(z,t)$, $\phi$ is the constant porosity of the soil,
taken here to be 0.25, $D_{0}D(S)$ and $K_{0}K(S)$ are the water
diffusivity and hydraulic conductivity respectively, with the
functions $D(S)$ and $K(S)$ given by
\begin{equation}\label{K(S)}
K(S) = S^{1/2}[1-(1-S^{1/m})^m]^2,
\end{equation}
\begin{equation}\label{D(S)}
D(S) = \frac{[1-(1-S^{1/m})^m]^2}{S^{1/m-1/2}(1-S^{1/m})^{m}},
\end{equation}
where $0<m<1$ (see \cite{genuchten80} and \cite{mualem76}).
The value of $m$ for the expanded-clay soil should be found experimentally
but, for later use in simulations and analysis of the model,
is taken to be $m=\frac{1}{2}$. Likewise the values of the
constants $K_{0}$, the conductivity for saturated soil, and
$D_{0}$, a representative value of diffusivity, should be
obtained empirically for particular roofing materials.
However, in the absence of good experiments, values as found
in \cite{roose04a} and \cite{roose04b} are assumed here.
Water uptake by the plant roots
is incorporated into the model through the last term in
(\ref{richards}) and is given (\cite{roose04a} and \cite{roose04b}) by
\begin{equation}
R = 2\pi a k_{r}l_{d} \left( p_{a} - p_{c}f(S) - p_{r} \right),
\end{equation}
where $k_{r}$ is the root's radial conductivity of water, $a$ is the root
radius, $l_{d}$ is the average number of roots per unit (horizontal)
area, $p_{a}$ is atmospheric
pressure, $p_{r}$ is an effective pressure in the roots (although
it can be negative), $p_{c}f(S)$ is the capillary pressure in the
soil, with $p_{c}$ another constant characterising the partly
saturated pellets, and
\begin{equation}\label{F(S)}
f(S)=(S^{-\frac{1}{m}}-1)^{1-m}.
\end{equation}
We take parameter values from Roose and Fowler \cite{roose04a} and
let $2\pi a k_{r}=7.85\times 10^{-16}$ m$^2$ s$^{-1}$ Pa$^{-1}$,
$l_{d}=5\times 10^3$ m$^{-2}$ and $p_{c}=10^4$ N m$^{-2}$. The
root pressure $p_{r}$ will be determined from conservation of
water within the root. Finally we must prescribe boundary
conditions at the top and bottom of the soil layer. At the soil
surface we take
\begin{equation}
D_{0}D(S)\frac{\partial S}{\partial
z}+K_{0}K(S)=Q_{in}(t)\quad\textrm{at}\quad z=L,
\end{equation}
where $Q_{in}$ is the rainfall rate averaged over the surface area
of the ground. We assume, in this unsaturated case, no outflow at
the base of the soil layer and set
\begin{equation}
D_{0}D(S)\frac{\partial S}{\partial
z}+K_{0}K(S)=0\quad\textrm{at}\quad z=0.
\end{equation}

\

We nondimensionalise the equations by scaling
\begin{equation}
z= L\hat{z},\quad p_{r}=|P|\hat{p_{r}},\quad t=
\frac{L}{K_{0}}\hat{t},\quad p=p_{a}+p_{c}\hat{p}, \quad R=2\pi a
k_{r} l_{d}|P|\hat{R}, \quad Q_{in}=Q_{typ}\hat{Q},
\end{equation}
where $P$ is the (negative) root pressure at the soil surface and
we set \\
$|P|=10^6$ N m$^{-2}$. The time scale used here is that for
flow though the soil layer under the action of gravity,
with saturation neither small nor close to one. The
dimensionless Richards' equation (\ref{richards}) then has the
form
\begin{equation}\label{dimensionless-richards-eqn}
\phi\frac{\partial
S}{\partial\hat{t}}=\frac{\partial}{\partial\hat{z}}\left(\delta
D(S)\frac{\partial S}{\partial \hat{z}}+K(S)\right)-\eta
(\theta-\varepsilon f(S)-\hat{p}_{r}),
\end{equation}
where
\begin{equation}\label{dimensionless parameters}
\delta=\frac{D_{0}}{LK_{0}}\approx 10^{-4},\quad \eta=\frac{2\pi a
k_{r} l_{d} |P| L}{K_{0}}\approx
4\times10^{-6},\quad\theta=\frac{p_{a}}{|P|}\approx
10^{-1},\quad\varepsilon=\frac{p_{c}}{|P|}\approx 10^{-2}.
\end{equation}
Roose and Fowler \cite{roose04a} give values of $D_{0}$ for
different soil types and we can reasonably take $D_{0}=10^{-6}$
m$^{2}$ s$^{-1}$. However, the value of $K_{0}$ is more difficult
to determine as it varies significantly with different soil types.
The parameter values in (\ref{dimensionless parameters}) are given
for $K_{0}=10^{-1}$ m s$^{-1}$. We note that $\eta\ll 1$
suggesting that water uptake by the roots is negligible over the
chosen time scale (of order 1~s). The dimensionless forms of the boundary
conditions are given by
\begin{equation}\label{BC1}
\delta D(S)\frac{\partial S}{\partial \hat{z}}+K(S)=\nu
\hat{Q}\quad\textrm{at}\quad \hat{z}=1,
\end{equation}
\begin{equation}\label{BC2}
\delta D(S)\frac{\partial S}{\partial
\hat{z}}+K(S)=0\quad\textrm{at}\quad \hat{z}=0,
\end{equation}
where
\begin{equation}
\nu=\frac{Q_{typ}}{K_{0}}\approx 3\times10^{-6},
\end{equation}
with $Q_{typ}$ taken to be some typical rainfall. We set
$Q_{typ}=3\times 10^{-7}$ m s$^{-1}$ for a ``wet day" in Ireland.
(This figure equates to about 2.6~cm in a day. Averaging
a monthly precipitation would give a substantially lower
figure.)\footnote{Rainfall of 8~cm in
30~minutes was
recorded at Eskdalemuir in southwest Scotland in 1953. Such a
figure would make $Q_{typ}$ over 100 times larger
but still keep $\nu$ small.} With this size of rainfall,
on order one times,
the saturation will generally be small, of order $\nu^{2/9}$ for
$m = 1/2$. This might suggest a rescaling of the saturation $S$,
but we delay such an approach until later. First, we address the 
possibility of the soil becoming fully saturated, the no-flux
condition at the base in (\ref{BC2}) indicating that the whole soil
layer would fill up on a dimensionless time scale of $O(\nu^{-1})$.
When the soil becomes saturated, however,
the model must change, and this allows for drainage through
the base as described below. (Note that the time scale for
$S$ to become locally order one near the base should depend
on $\delta$ as well as on $\nu$.)


\subsection{The Saturated Region}
When the soil starts to become fully saturated ($S=1$) at
$z=0$, we assume that a moving boundary
forms between the fully saturated soil below and the partially
saturated soil above. This boundary lies at $z=h(t)$
and the soil saturation is identically one for
$z\in[0,h]$.  For saturated soil, water flux is given by
Darcy's law,
\begin{equation}\label{fluxB}
q = -  K_0 \left( \frac{\partial p}{\partial z} + \rho g \right),
\end{equation}
instead of (\ref{fluxA}), and
our governing equation in this lower region
can now be written in the
(dimensionless) form
\begin{equation}\label{pr-ode}
\frac{\partial}{\partial
\hat{z}}\left(1+\frac{1}{\gamma}\frac{\partial \hat{p}}{\partial
\hat{z}}\right)-\eta\left(\theta+\varepsilon\hat{p}-\hat{p}_{r}\right)=0,
\end{equation}
where
\begin{equation}
\gamma=\frac{\rho gL}{p_{c}}\approx 10^{-1}.
\end{equation}
The flux through the membrane at $z=0$ is prescribed to occur at a
rate proportional to the pressure difference across it: $Q_{mem} =
\kappa (p-p_a)$ dimensionally, where $p_a$ is the atmospheric
pressure in the drainage layer beneath, $p$ is the pressure at
$z=0$, and $\kappa \approx 10^{-5} \ \textrm{m}\ \textrm{s}^{-1} \
\textrm{Pa}^{-1}$ (determined experimentally in the next
sub-section). This gives the dimensionless condition
\begin{equation} \label{090706.3}
 1 +\frac{1}{\gamma} \frac{\partial \hat{p}}{\partial \hat{z}}
= \alpha \hat{p} \quad \textrm{at} \quad \hat{z}=0,
\end{equation}
where $\alpha = \frac{\kappa p_c}{K_0}\approx 1$. At the
saturation front $\hat{z}={\hat h}({\hat t})\equiv
\frac{h(t)}{L}$, $S = 1$, $\hat{p}=0$ (atmospheric), and continuity of
fluid flux requires
\begin{equation} \label{090706.4}
\lim_{\hat{z}\to\hat{h}+} \left[ K(S) + \delta 
D(S)\frac{\partial S}{\partial \hat{z}} \right] =
\lim_{\hat{z}\to\hat{h}-} \left[ 1 +
\frac{1}{\gamma} \frac{\partial \hat{p}} {\partial \hat{z}} \right] .
\end{equation}
Neglecting the $\eta$ term in (\ref{pr-ode}) for this saturated
region, and using $\hat{p} = 0$ at $\hat{z}=\hat{h}$ along with
(\ref{090706.3}), gives
\begin{equation}
\hat{p}(\hat{z},\hat{t}) = \frac{\gamma
(\hat{h}(\hat{t})-\hat{z})}{1+\alpha \gamma \hat{h}(\hat{t})},
\end{equation}
so that (\ref{090706.4}) becomes
\begin{equation} \label{090706.7}
\lim_{\hat{z}\to\hat{h}+} \left(
K(S) + \delta D(S)\frac{\partial S}{\partial \hat{z}} \right) =
\frac{\alpha \gamma \hat{h}}{1 + \alpha
 \gamma \hat{h}}
\end{equation}
and then, on using $K(1) = 1$,
\begin{equation} \label{bottom.new}
- \lim_{\hat{z}\to\hat{h}+} \left( \delta D(S)
\frac{\partial S}{\partial \hat{z}} \right) =
\frac 1{1 + \alpha
 \gamma \hat{h}} \, .
\end{equation}
In principle equation (\ref{090706.7}) and boundary condition
$S=1$ at $\hat{z}=\hat{h}(\hat{t})$ determine $\hat{h}$ in terms
of the flux from the unsaturated region. However, we can simplify
things if we notice from (\ref{BC1}) that the dimensionless flux
will in general be small, of order $\nu$ (due to the rainfall). If
this is the case, then the value of $\hat{h}$ required to satisfy
(\ref{090706.7}) will be small. Physically, this is because, for
the typical size of fluid flux considered, the pressure required
to force it through the membrane according to (\ref{090706.3}) is
provided by the hydrostatic head of a very thin layer of water
(dimensionally, $h$ is calculated to be much less than $1\
\textrm{mm}$).

Thus if a saturated region is created at the bottom of the soil
layer, it will quickly grow to a depth which is sufficient to
drain exactly the same amount of water through the membrane as is
arriving from the unsaturated region above. Provided this depth is
substantially less than the depth of the soil, the saturated
region can be `collapsed' (mathematically) onto the line
$\hat{z}=0$, and the boundary condition applied to the problem in
the unsaturated zone for some of the numerical solutions of
sub-section 2.5 is then simply
\begin{equation} \label{S10}
S = 1 \quad \textrm{at} \quad \hat{z}=0 \, .
\end{equation}
After computing the solution $S(\zhat,\that)$
of the problem with the simplified boundary condition (\ref{S10}),
we can evaluate the limit in (\ref{bottom.new}), and hence estimate
the small non-zero depth $h(\zhat,\that)$.

Note that with this model, even with the $\eta$ term restored
in the saturated layer, once the layer forms, there is no mechanism
by which it will entirely disappear:

Starting with a completely unsaturated roof, so that (\ref{BC2})
is initially imposed at the base, if the roof attains saturation
at some dimensionless time ${\hat{t}}_s$, the base condition
(\ref{S10}) holds for all later times $\hat{t} > {\hat{t}}_s$.
\footnote{Alternatively, the bottom
condition might be specified in linear complementary form
$(1-S)q=0$ with $1-S\ge 0$ (for no super-saturation) and $q\le 0$
(for downward flux).}



\subsection{Experimental measurement of $\kappa$} 
\begin{figure}
\begin{center}
\scalebox{0.6}{\includegraphics{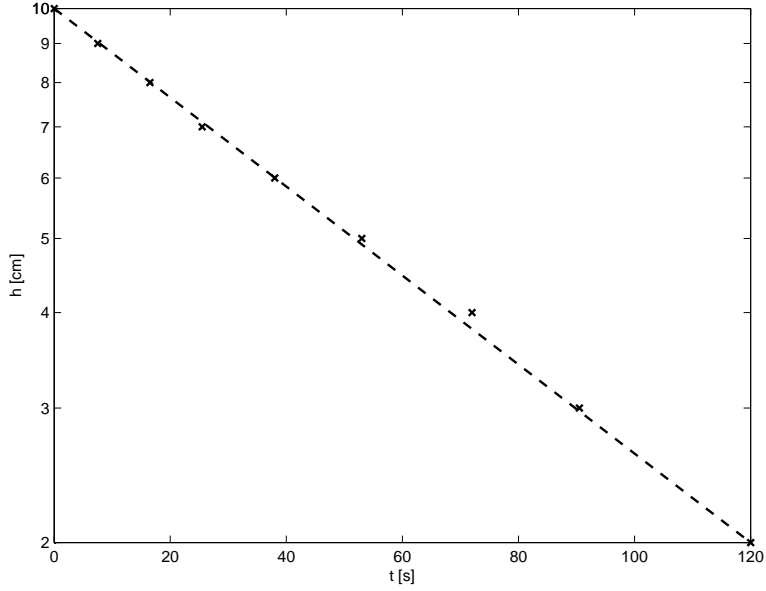}}
\end{center}
\caption{Experimental measurements of $h$ against $t$.}%
\label{experiment}%
\end{figure}

The value of
$\kappa$ was deduced from a simple experiment, which involved
puncturing a $2\ \textrm{mm}$ diameter hole in a plastic bottle,
made with material similar to that of the drainage
membrane (this is normally made from high-density polyethylene).
The rate of drainage through the hole driven
by the hydraulic head in the bottle was measured, and used to
determine the coefficient of proportionality between pressure
difference across the membrane $\Delta p$ and the water flux
through it $q$. Writing
\begin{equation}
q = k \Delta p,
\end{equation}
where $\Delta p = \rho g h$, the water depth in the bottle, $h$,
satisfies the equation
\begin{equation}
A_{bottle} \frac{d h}{dt} = - k \rho g h,
\end{equation}
where $A_{bottle}$ is the cross-sectional area of the bottle. Thus
\begin{equation}
\log h = - \frac{ k \rho g }{A_{bottle}} t.
\end{equation}
Measurements of $h$ against $t$ made during the experiment are in
Fig. \ref{experiment}, and the best fit value of the time constant
$t_c = A_{bottle} / k \rho g$ was $74$ seconds.  The flux through
an individual hole can be converted into an average velocity
through a membrane, using the area of the membrane $A_{membrane}$
that is drained by each hole. Thus
\begin{equation}
\bar{u} = \kappa \Delta p, \qquad \kappa = \frac{A_{bottle}}
   {A_{membrane} \rho g t_c}.
\end{equation}
Taking $A_{membrane} = \pi\ \textrm{cm}^2$, and using the
cross-sectional area of the bottle $A_{bottle} = 25\
\textrm{cm}^2$, $\rho = 10^3 \ \textrm{kg}\ \textrm{m}^{-3}$, and
$g = 10\ \textrm{m}\ \textrm{s}^{-2}$, gives $\kappa \approx
10^{-5} \ \textrm{m}\ \textrm {s}^{-1}\ \textrm{Pa}^{-1}$.


\subsection{The Root Pressure}
To determine the root pressure $p_{r}$ in equation
(\ref{dimensionless-richards-eqn}), we assume that the root
extends through the full thickness of the soil layer of depth $L$.
Conservation of water inside the root yields
\begin{equation}
k_{z}\frac{d^{2}p_{r}}{d z^{2}}+2\pi a
k_{r}\left(p_{a}-p_{c}f(S)-p_{r}\right)=0,
\end{equation}
where $k_{z}=10^{-14}$ m$^{6}$ s$^{-1}$ N$^{-1}$ is the root axial
conductivity and $f(S)$ is defined in equation (\ref{F(S)}). Zero
axial flux at the root tip implies
\begin{equation}
\frac{dp_{r}}{dz}+\rho g=0\quad\textrm{at}\quad z=0.
\end{equation}
In addition we prescribe a driving pressure, $P$, at the root base
yielding
\begin{equation}
p_{r} = p_{a} + P \quad \textrm{at} \quad z = L .
\end{equation}
In dimensionless form the root pressure will satisfy
\begin{equation}\label{prODE}
\frac{d^{2}\hat{p}_{r}}{d
\hat{z}^{2}}+\tau\left(\theta-\varepsilon
f(S)-\hat{p}_{r}\right) = 0 \quad \mbox{ in } \, 0 < \zhat < 1 \, ,
\end{equation}
subject to
\begin{equation}
\frac{d\hat{p}_{r}}{d\hat{z}}=-\varepsilon\gamma\quad\textrm{at}\quad
\hat{z}=0,
\end{equation}
\begin{equation}
\hat{p}_{r}=\theta-1\quad\textrm{at}\quad \hat{z}=1,
\end{equation}
where
\begin{equation}
\tau=\frac{2\pi a k_{r}L^{2}}{k_{z}}\approx 10^{-3}.
\end{equation}
The parameters $\tau\ll 1$ and $\varepsilon\gamma\ll 1$ which
implies $\displaystyle \frac{d^{2}\hat{p}_{r}}{d\hat{z}^{2}}
\approx 0$ subject
to $\frac{d\hat{p}_{r}}{d\hat{z}}=0$ on $\hat{z}=0$. 
Note that having $\tau\ll 1$ means that varying saturation in the
soil has negligible effect on the root pressure. The 
dimensionless root pressure is thus given by
\begin{equation}\label{pr}
\hat{p}_{r} = \theta - 1 \quad \mbox{ for } \, 0 \le \zhat \le 1 \, .
\end{equation}
\medskip
The complete model is now given by
(\ref{dimensionless-richards-eqn}), with the definitions
(\ref{K(S)}), (\ref{D(S)}), (\ref{F(S)}) and (\ref{pr}), with
boundary condition (\ref{BC1}) at $\hat{z}=1$ and (\ref{BC2}) if
$S < 1$, or (\ref{S10}) otherwise. An
initial condition is also needed.

The diffusion term which has $\delta$ as a factor
in (\ref{dimensionless-richards-eqn}) is small,
so the equation is essentially a first-order non-linear wave
equation; the boundary condition (rainfall) is transmitted
downwards as a wave. If rain starts suddenly, there is a sharp
jump in saturation that propagates quickly down
to the bottom of the soil; if the rain
stops suddenly then, in the $z$ -- $t$ plane,
the solution is described by a classical expansion fan.


\subsection{Numerical Solutions}
\begin{figure}[htp]
\begin{center}
\scalebox{0.65}[0.65]{\includegraphics{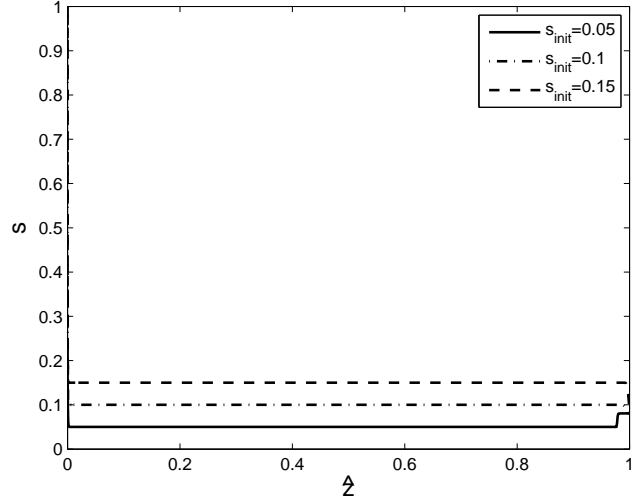}}
\end{center}
\caption{$\emph{S}$ {\it vs.} $\hat{z}$ for three different initial
conditions ($S_{init}=0.05,0.1,0.15$) at either  dimensionless
time 100 ($S_{init}=0.05$) or when $S$ reaches 1 at $\hat{z}=0$
($S_{init}=0.1,0.15$). Parameter values are $m=1/2$,
$\delta=10^{-4}$, $\nu=3\times 10^{-6}$.
The top condition has $\Qhat = 1$. Note a sudden increase
to $S=1$ for small $\zhat$.} %
\label{fig1}%
\end{figure}

\begin{figure}[htp]
\begin{center}
\scalebox{0.65}[0.65]{\includegraphics{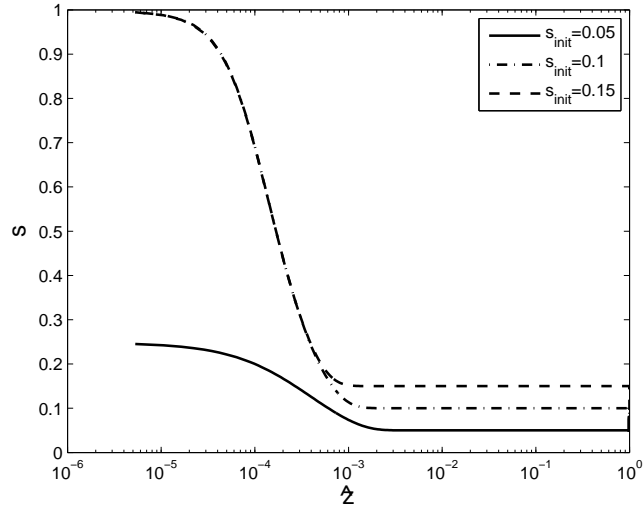}}
\end{center}
\caption{A semi-log plot of $\emph{S}$ {\it vs.} $\hat{z}$ for three
different initial conditions ($S_{init}=0.05,0.1,0.15$) at either
dimensionless time 100 ($S_{init}=0.05$) or when $S$ reaches 1 at
$\hat{z}=0$ ($S_{init}=0.1,0.15$). Parameter values are $m=1/2$,
$\delta=10^{-4}$, $\nu=3\times 10^{-6}$.}%
\label{fig2}%
\end{figure}

\begin{figure}[htp]
\begin{center}
\scalebox{0.65}[0.65]{\includegraphics{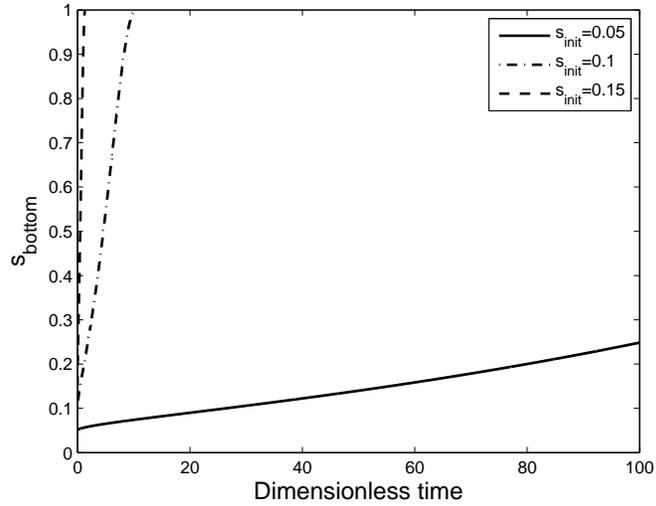}}
\end{center}
\caption{$\emph{S}_{bottom}$ {\it vs.} dimensionless time for three
different initial conditions ($S_{init}=0.05,0.1,0.15$). Parameter
values are $m=1/2$,
$\delta=10^{-4}$, $\nu=3\times 10^{-6}$.}%
\label{fig3}%
\end{figure}

\begin{figure}[htp]
\begin{center}
\scalebox{0.65}[0.65]{\includegraphics{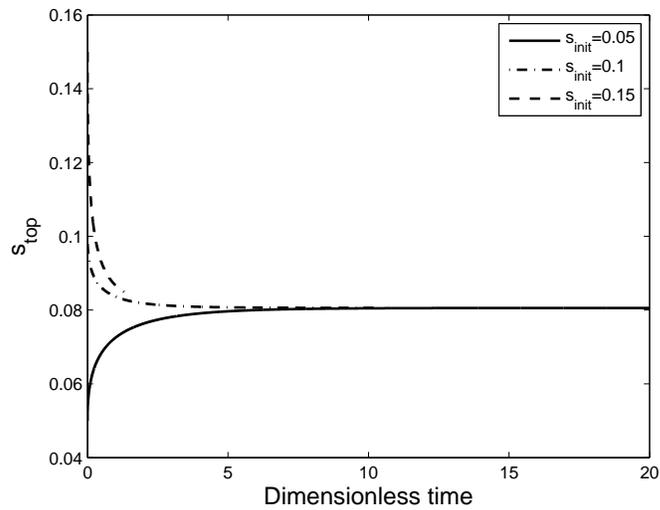}}
\end{center}
\caption{$\emph{S}_{top}$ {\it vs.} dimensionless time for three
different initial conditions ($S_{init}=0.05,0.1,0.15$). Parameter
values are $m=1/2$,
$\delta=10^{-4}$, $\nu=3\times 10^{-6}$.
The top condition has $\Qhat = 1$.}%
\label{fig4}%
\end{figure}

The governing equation for the unsaturated region
(\ref{dimensionless-richards-eqn}) was solved numerically
subject to boundary
conditions (\ref{BC1}), with $\hat{Q} = 1$, and (\ref{BC2}).
In these first simulations, a finite element method was
used with 385 elements
and significant refinement near $\zhat = 0$ and near $\zhat = 1$.
As a first approach the
$\eta$ term in (\ref{dimensionless-richards-eqn}) is neglected so
that we are just considering drainage of the soil layer under
gravity. The initial saturation was taken to be uniform throughout
the soil layer. Three different initial values of the saturation
$S_{init}=0.05,0.1,0.15$ were considered. The profiles obtained
for $S$ in all the cases, when the computations were
stopped, are shown in
Fig.~\ref{fig1}; a corresponding semi-log plot is shown in
Fig.~\ref{fig2}, in order to demonstrate the boundary layer of
thickness $\delta$ in $S$ at $\hat{z}=0$ that is predicted
by comparing the two transport terms in
(\ref{dimensionless-richards-eqn}),
and which is captured by the numerical
solution, but which is not visible in Fig.~\ref{fig1}. For
$S_{init}=0.1$ and 0.15, computations were stopped when the value
of  $S$ at $\hat{z}=0,$ $S_{bottom},$ reached 1; for
$S_{init}=0.05,$ $S_{bottom}$ is still far from 1, even for the
value of dimensionless time (100) shown here. The time evolution
of $S_{bottom}$ is shown in Fig.~\ref{fig3}, \ while that for $S$
at $\hat{z}=1,$ $S_{top},$ is shown in Fig.~\ref{fig4}.

Thus, the results suggest an appreciable difference in the time at
which complete saturation is achieved at the bottom of the soil
when $S_{init}$ is increased from $0.05$ to $0.1$. The effect of
the rainfall boundary condition (\ref{BC1}) has (by the end of the
simulations) only affected the tiny region at the right of Fig. 3,
where there is the beginning of a shock front propagating
downwards from $\hat{z}=1$; since $\delta$ has been taken to be
very small, the shock looks very sharp, and the values on either
side of it are the initial condition (below, or left, of the
shock), and the value given by $K(S) = \nu \hat{Q}$ (above, or
right, of the shock - this value is expectedly independent of the
initial condition, as shown in Fig. \ref{fig4}).\\

\

The complete problem, with a small saturated region allowed for by
using boundary condition (\ref{S10}), and with $\eta\neq 0$,
was also solved by discretising in space and solving with the
method of lines using ode15s in Matlab. Rather than have a mesh
refinement as employed earlier to cope with the stiffness produced
by the small value of $\delta$, the value of this parameter is
now taken to be artificially large, $\delta = 10^{-2}$.
We use a larger value of $\delta$ partly so as to avoid
having to use a variable grid and partly so as to make
the diffusive transition layers more clear visible in the solutions.
Since the value is still small, using the larger value
does not affect the overall dynamics -- it simply exaggerates
the width of the diffusive layers.
To apply the switch in
boundary conditions smoothly, the condition
\begin{equation}\label{q}
q_0 = q_1 e^{-1000(1-S)},
\end{equation}
was applied for the flux at the bottom node $q_0$ in terms of the
flux at the node above $q_1$; thus when $S$ is close to $1$ this
becomes $\partial \hat{q} /
\partial \hat{z} = 0$, and when $S$ is less than $1$ it becomes $q_0 = 0$.
The diffusion coefficient is infinite when $S=1$, but this does
not cause any issues in the numerics, possibly because the above
boundary condition ensures $S$ never quite reaches $1$.

\begin{figure}
\centering
\includegraphics[width=\textwidth]{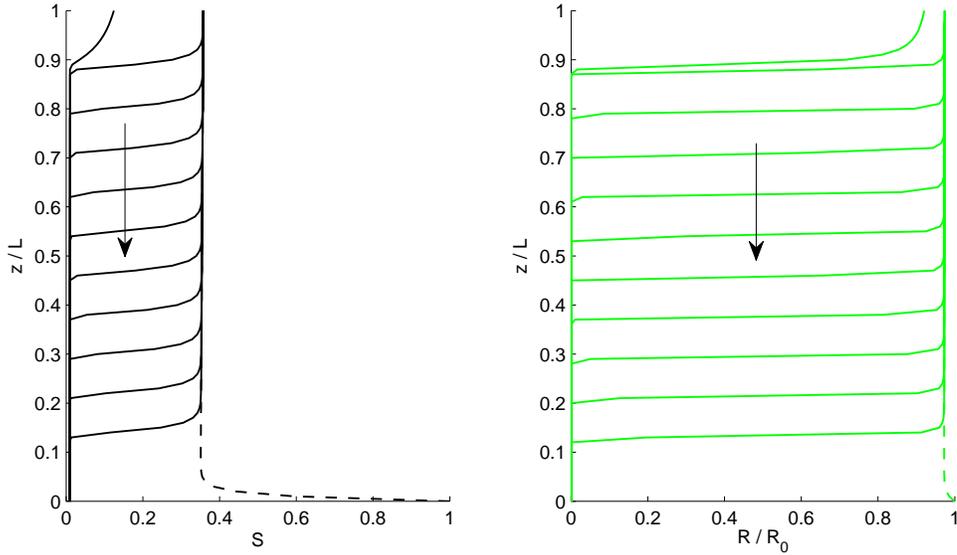}
\caption{\label{greenroof_1} Profiles of saturation and root
uptake at time intervals of $1$ (in the dimensionless units); the
arrow shows the direction of increasing time. This is the result
of a sudden increase in rainfall to $\hat{Q} = 10$, from the
steady state when $\hat{Q} = 0.1$, and the dashed line shows the
steady state that results.  Parameter values are $m=1/2$,
$\delta=10^{-2}$, $\eta = 4\times 10^{-4}$, $\nu=3\times 10^{-4}$,
$\varepsilon = 10^{-2}$, $\gamma = 10^{-1}$.}
\end{figure}

This seems to allow for steady states when rainfall is constant;
if there is more rainfall than is taken up by the roots, the
saturation at the bottom is $1$ and there is a boundary layer of
width $\delta$ in which it adjusts to the value as
determined by $K(S) \approx \nu \hat{Q} - \eta \int_0^1 \Rhat \,\dd\zhat$
(Fig.~\ref{greenroof_1}).  If there is less rainfall than is taken up by
the roots, the saturation at the bottom decreases almost to $0$.

\begin{figure}
\centering
\includegraphics[width=\textwidth]{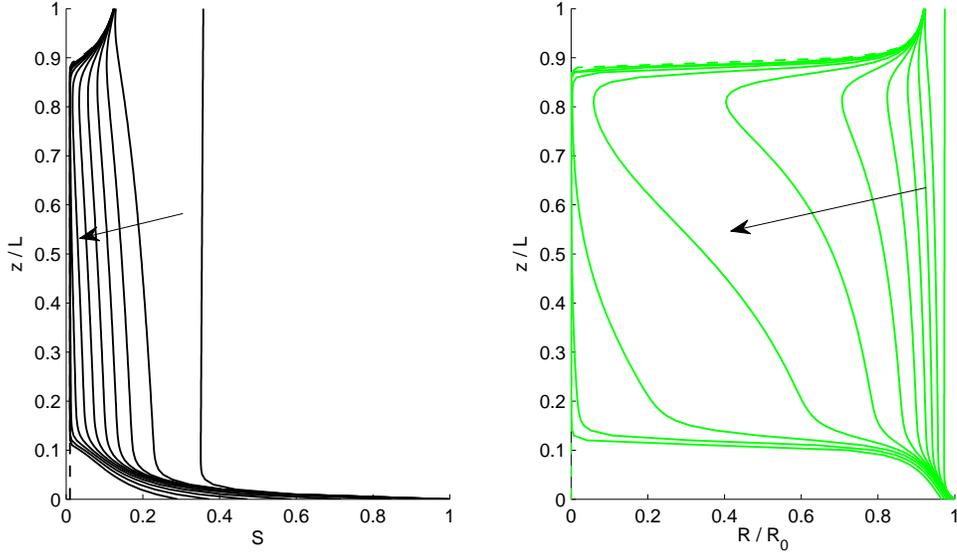}
\caption{\label{greenroof_2} Profiles of saturation and root
uptake at time intervals of $10$ (in the dimensionless units); the
arrow shows the direction of increasing time. This is the result
of a sudden decrease in rainfall from $\hat{Q} = 10$, to $\hat{Q}
= 0.1$. Parameter values are $m=1/2$, $\delta=10^{-2}$, $\eta =
4\times 10^{-4}$, $\nu=3\times 10^{-4}$, $\varepsilon = 10^{-2}$,
$\gamma = 10^{-1}$.}
\end{figure}

Fig.~\ref{greenroof_1} shows the result of a sudden increase in
rainfall from $\hat{Q} = 0.1$ to $\hat{Q}=10$, which shows the
initial shock front travelling down into the soil and the eventual
steady state. The saturation at the bottom does not increase
towards $1$ until the shock front arrives there.
Fig.~\ref{greenroof_2} shows the result of a sudden decrease back to
$\hat{Q}=0.1$. Note that the time intervals shown are longer.
Most of the apparent changes occur over a time scale suggested
by following characteristics (neglecting the diffusion term)
from $\zhat =1$ where the saturation is given by $K(S) = \nu \hat{Q}$,
say $S = S_1$. Along such a characteristic, $S$ is given by
\[
\phi\od S\that = - \eta (\theta \epsilon f(S) - \phat_r)
\sim \eta \left( \frac\epsilon S - 1 \right)
\]
for $S$ small, and the (dimensionless) time scale is of order
$S_1/(\phi\eta) \approx 100$ for this particular problem.
(This time scale may be associated with an expansion fan localised
near $\zhat = 1$.) For this case, there appears to be a more
substantial boundary layer, possibly of width $\delta^{1/2}$,
near $\zhat = 0$.

\

The simple model presented in this section suggests that we can
generally expect the soil to be partially saturated throughout
most of its depth, with a small saturated layer at the base
facilitating drainage through the underlying membrane.
Even with quite large rainfall, the drainage is apparently
sufficient to evacuate the water without the soil becoming
fully flooded.  This is of course dependent on the permeability
of the membrane, which may vary considerably and may also
decrease with time due to clogging; but given the values assumed
here we may conclude that full saturation of the soil layer
is unlikely. On the other hand the model suggests the opposite
problem of having long periods of drought when there is no rainfall.
We therefore turn to some alternative two-porosity models that
could give longer-term water storage.


\section{Two-Porosity Models}
The expanded clay pellets used in green roof construction are
quite large but contain lots of pore space. The difference in pore
sizes between these and the inter-pellet space means water can be
drawn into the pellets and retained there for longer than it would
otherwise remain in the soil. Thus a two-porosity model would seem
appropriate.


\subsection{A Model with Slow Saturation}

This is an outline of a ``box" or ``lumped" model for water
storage in the macro-pores between soil particles, which have
saturation $S$, and in the micro-pores within the particles, which
have saturation $S_P$.  Transport of water into or
out of the particles is given by a rate constant $\lambda$ times
the saturation difference $S - S_P$ (the penultimate term in
(\ref{090706.1}) and the right-hand side of (\ref{090706.2}),
below).\footnote{A variant of this model
might assume that water transfer into the particles occurs at a
rate proportional to the \emph{pressure} difference $p_{cP}
f(S_P)- p_c f(S)$; since the capillary pressure in the micropores
would be larger than in the macropores ($p_{cP} > p_c$), this
would cause more water to be transferred into the micropores, and
a larger supply would be maintained there for the roots to take
up.} The roots do not penetrate into individual particles so
provide a sink term $R$ only from the macro-pores. This root
uptake $R(S)$ in (\ref{richards}) is primarily due to
the large negative pressure in
the root system, but as saturation decreases a large capillary
pressure acts to counteract this; thus $R(S)$ is roughly constant
for $S$ close to $1$ but decreases at small $S$ (as in the model
above).

The following equations are dimensionless, and the time scale has
been chosen to be that due to uptake by the roots (the time scale
differs from that used previously by a factor $\eta$,
so that now $t = t_0\that$ with $t_0 = 1/(2\pi a k_r l_d |P|)
\approx 2.5\times10^{5}s\approx$ 3~days). Drainage
from the volume of soil is supposed to occur due to gravity at a
rate $K(S)$, and occurs on a time scale $\eta$ compared to the
uptake by the roots (see above). Rainfall provides a source which
is scaled to be the same size as the gravity drainage (note this
is different to above -- the scale for the rainfall here is large
and is intended to represent the size of heavy showers; the
dimensionless $r(\hat{t})$ will be $0$ most of the time, when it
is not raining, and $O(1)$ when it is raining heavily).

\begin{equation} \label{090706.1}
\phi\od S\that = \frac{1}{\eta} r(\hat{t}) -
\frac{1}{\eta} K(S) - \lambda (S - S_P) - R(S),
\end{equation}
\begin{equation} \label{090706.2}
(1-\phi)\phi_{P} \od{S_P}\that = \lambda (S - S_P),
\end{equation}
where $r=\nu\hat{Q}$, $\phi_{P}$ is the porosity of the pellets,
 $\lambda > 0$ is a transport constant and
\begin{equation}
K(S) = S^{1/2}[1-(1-S^{2})^{1/2}]^2,
\end{equation}
which comes from equation (\ref{K(S)}) with $m=\frac{1}{2}$ and
\begin{equation}\label{R}
R(S) = 1 - \varepsilon \frac{(1-S^{2})^{1/2}}{S}.
\end{equation}

The use of $K(S)$ for the gravity drainage in equation
(\ref{090706.1}) is motivated by the fact that the water flow
in Section~2 is essentially determined by this hydraulic
conductivity (since $\delta$ is small). The time
scale for water to diffuse into individual particles is estimated
using their dimensions $L_P \sim 1\ \textrm{cm}$ and a diffusion
coefficient $D_P \sim 10^{-9}\ \textrm{m}^2\ \textrm{s}^{-1}$.
$L_P^2/D_P$ is comparable to the time scale for uptake by the
roots ($\sim 10^{5} \ \textrm{s}$), so the parameter $\lambda$ is
order $1$.  In equation (\ref{090706.1}) $\eta$ is very small and
in equation (\ref{R}) $\varepsilon$ is also small, and we consider
especially the distinguished case $\varepsilon$
of order $\eta^{2/9}$, see (\ref{Rbalance1}) below.

\begin{figure}
\centering
\includegraphics[width=\textwidth]{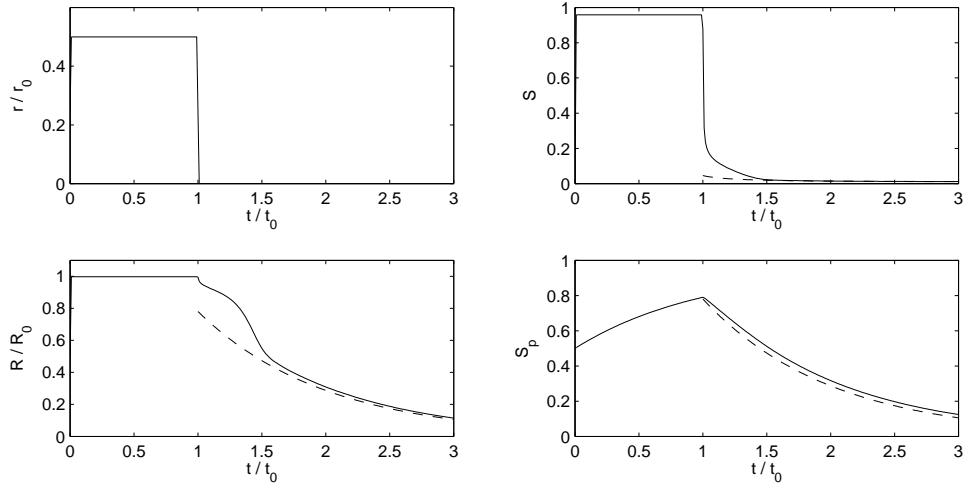}
\caption{\label{greenroof3} Solutions for macro-scale and
particle-scale saturations $S$ and $S_P$, and root uptake $R$, as
a result of rainfall $r(t)$ which represents a large rain shower.
Dashed lines show the limiting behaviour.  Parameter values are
$\eta = 10^{-4}$, $\lambda = 1$, $\varepsilon = 10^{-2}$.}
\end{figure}

The behaviour of solutions to this model is quite straightforward,
and an example solution for a large rain storm followed by dry
weather is in Fig.~\ref{greenroof3}. When it is raining, $r$ is
order $1$, and on a fast time scale, $\hat{t} \sim O(\eta)$, the
saturation $S$ relaxes towards the equilibrium given by $K(S) =
r(\hat{t})$. This causes water to then transfer into the particles
on an $O(1)$ time scale according to (\ref{090706.2}).
When it stops raining $r
= 0$, and the saturation $S$ decreases quickly due to gravity
drainage on an $O(\eta)$ time scale. In this fast regime, 
(\ref{090706.1}) is approximately
\begin{equation}
\phi\od S\that \sim - \frac{1}{\eta} K(S),
\end{equation}
where $K(S) \sim \frac{1}{4}S^{\frac{9}{2}}$ for $S$ small.
This suggests
$S$ tends towards $0$ as
\begin{equation}
S \sim \left( \frac{8\phi \eta}{7 \hat{t}} \right)^{2/7}.
\end{equation}
Looking to balance the $\dd S/\dd \that$, $K(S)/\eta \sim
S^{\frac{9}{2}}/(4\eta)$ (for $S$ small), and $\lambda(S - S_P)
\sim - \lambda S_P$ (for $S$ small) terms in (\ref{090706.1}),
we then take $\that = \eta^{2/9} \tilde{t}$ and $S = \eta^{2/9}
\tilde{S}$. A complete balance from the final term,
\begin{equation} \label{Rbalance1}
R(S) \sim 1 - \varepsilon/( \eta^{2/9} \tilde{S})
\end{equation}
is achieved on taking
\begin{equation} \label{Rbalance2}
 \varepsilon = \eta^{2/9} \tilde{\varepsilon} .
\end{equation}
In this intermediate regime, (\ref{090706.1}) then reduces to
\begin{equation} \label{inter1}
\phi\od {\tilde{S}}{\tilde{t}} \sim - \frac 14 {\tilde{S}}^{9/2} +
\lambda S_P - 1 + \frac{\tilde{\varepsilon}}{\tilde{S}}
\end{equation}
while (\ref{090706.2}) becomes simply, to leading order,
\begin{equation} \label{inter2}
\od {S_P}{\tilde{t}} = 0 .
\end{equation}
On the $O(1)$ time scale, $S$ continues to be order $\eta^{2/9}$
and can be regarded as quasi-stationary,
with (\ref{090706.1}) (or (\ref{inter1})) being replaced by
\begin{equation} \label{order1.1}
\frac{K(S)}\eta + R(S) \sim \frac 14 {\tilde{S}}^{9/2} + 1 -
\frac{\tilde{\varepsilon}}{\tilde{S}} \sim
\lambda S_P
\end{equation}
while $S_P$ now reduces, with (\ref{090706.2}) (or (\ref{inter2}))
being replaced by
\begin{equation}
(1-\phi)\phi_{P}\od{S_P}\that \sim - \lambda S_P .
\end{equation}
Thus, ignoring the small terms, $S_p$ decays
exponentially and the water coming out into the macropores is
immediately either taken up by the roots or lost by drainage:
\begin{equation} \label{SP-last}
R + \frac K\eta = \lambda S_P = -(1-\phi)\phi_{P}dS_P/d\hat{t} ,
\end{equation}
In the example shown in Fig.~\ref{greenroof3}, $\tilde{\varepsilon}$
is rather small (approximately $0.1$) while $\lambda S_P$ is
significantly less than one in this time regime. Equation
(\ref{order1.1}) then indicates that $\tilde{S}$ is small
and the water being lost by drainage is negligible; in this
case the water coming out of the micropores is
immediately taken up by the roots.

A longer-time regime will
apply for $S_P$ sufficiently small but this is not considered here.

\

In conclusion, root uptake is maintained for a much longer
period (as it decreases slowly with time according to
(\ref{SP-last})) after it ceases to rain. This contrasts with
the case with no
micropores, when $S$ decreases rapidly towards $0$ (the time scale
being a factor $\eta$ shorter).


\subsection{A Model for Fast Saturation}
Assuming instead fast saturation of the pellets,
so $S_P = \He(S)$,
the intra-pellet water content is given by
\begin{equation}
\mbox{water density in individual pellets} =
\phi_P S_P = \phi_P \He (S) \, , \label{twosats}
\end{equation}
where H denotes the Heaviside function, $S_P$ denotes the
saturation of the individual pellets, $\phi_P$ is the porosity of an
individual pellet, and $S$ is the saturation of the inter-pellet
pores. The required short time scale can arise from high capillary
pressures associated with the very small pores within the pellets.

Taking now $\phi = \frac 14$ to be the total proportion of space
occupied by air and water within the soil, and $\phi_P = \frac 15$,
then the inter-pellet porosity is $\varphi = \frac  1{16}$ (given by
$\varphi + \frac 15 (1 - \varphi) = \frac 14$). The total water
content is now inter-pellet water content (porosity $\times$
inter-pellet saturation), $\varphi S$, plus that of the pellets
(the volume fraction occupied by the pellets $\times$ their
porosity $\times$ their saturation), $(1 - \varphi) \phi_P S_P$, 
\[
= \frac 1{16} S + \frac{15}{16} \times \frac 15 \He (S)
= \frac 1{16} (S + 3\He (S)) .
\]
The water flux, $q$, and rate of uptake of water by the roots,
$R$, are assumed to depend on the inter-pellet saturation $S$
in the same way as earlier. Equation
(\ref{dimensionless-richards-eqn}) can then be replaced by
\begin{equation}
\frac 1{16}\pd{}\that (S + 3\He (S)) = \pd{}\zhat \left( K(S) +
\delta D(S) \pd S\zhat \right) - \eta\Rhat \, , \label{twosatspde}
\end{equation}
with $\Rhat \sim 1$, from (\ref{dimensionless-richards-eqn}) and
(\ref{pr}). (Equation (\ref{twosatspde}) might be better written
in terms of the total water content, $S_T = \frac 1{16}(S + 3\He
(S))$, so that $S$ on the right-hand side is replaced by $S(S_T) =
0$ for $0 \le S_T \le \frac 3{16}$, $S(S_T) = 16(S_T - \frac
3{16})$ for $\frac 3{16} \le S_T \le \frac 14$.)\\

Where the pellets are saturated, $S>0$ and $\He (S) = 1$, the
equations are as in Section 2. Here, for simplicity, an initially
dry soil is considered, so that at $\that = 0$, $S \equiv S_P
\equiv 0$. For $\that > 0$, a region $\What (\that ) < \zhat < 1$
has become wet:
\begin{equation}
S = \He (S) = 0 \, \mbox{ in } \, 0 < \zhat < \What \, , \quad S >
0 \, \mbox{ and } \, \He (S) = 1 \, \mbox{ in } \, \What < \zhat <
1 \, . \label{twosatswet}
\end{equation}
To obtain an order-one sized wet region, the relevant time scale
must be that for the rainfall (days). Hence time has to be
rescaled by
\begin{equation}
\that = \ttil /\nu \, . \label{twosatstime}
\end{equation}
Note that this time scale is similar to that for the up-take of
water by the plants' roots. It is also appropriate, from the top
boundary condition, to rescale the saturation:
\begin{equation}
S = \nu^{2/9}\Stil \, , \label{twosatssat}
\end{equation}
where, since we have assumed that $m = \frac 12$, $K(S) \sim \frac
14 S^{9/2}$ and $D(S) \sim \frac 14 S^{5/2}$ for small $S$.

Neglecting the time-derivative term (now effectively of order
$\nu^{2/9}$), the partial differential equation (\ref{twosatspde})
becomes
\begin{equation}
\frac 14 \pd{}\zhat \left( \Stil^{9/2} + \deltatil \Stil^{5/2} \pd
\Stil\zhat \right) = \etatil\Rhat \, . \label{twosatsPDE}
\end{equation}
Here $\etatil = \eta/\nu \approx \frac 13$ and $\deltatil =
\delta/\nu^{2/9} \approx \frac 1{600}$, using the values of
Section 2. Although the value of $\deltatil$ is small here,
because of the uncertainty in the values of the physical
parameters describing water transport through the soil, it could
conceivably be of order one and it is therefore retained in
(\ref{twosatsPDE}), for the present. The $\deltatil$
term should also be kept as it contains the highest derivative
in the equation, just as the diffusion term was retained in
Section 2.

The differential equation is subject to the top boundary condition
\begin{equation}
\frac 14 \left( \Stil^{9/2} + \deltatil \Stil^{5/2} \pd \Stil\zhat
\right) = \Qhat_{in} \, \mbox{ at } \, \zhat = 1
\label{twosatsbc1}
\end{equation}
and, assuming that the diffusive, $\deltatil$, term is retained, to a
lower boundary condition
\begin{equation}
\Stil = 0 \, \mbox{ at } \, \zhat = \What (\ttil ) \, .
\label{twosatsbc2}
\end{equation}

Finally, to fix the position of the free boundary $\zhat = \What
(\that )$ between dry and wet soil, conservation of mass of water
at this point, where $S_P$ jumps from 0 to $\phi_P$, leads to
\begin{equation}
\od\What\ttil = - \frac 43 \left( \Stil^{9/2} + \deltatil
\Stil^{5/2} \pd \Stil\zhat \right) \, \mbox{ at } \, \zhat = \What
(\ttil ) \, . \label{twosatsfbc}
\end{equation}
(Since, for $\deltatil > 0$, $\Stil = 0$ at this point, the second
term on the right-hand side should then be interpreted as
$\displaystyle \deltatil \lim_{\zhat \to \What} \left\{
\Stil^{5/2} \pd \Stil\zhat \right\}$.)

Of course, if the pellets were already partially saturated,
(\ref{twosatsfbc}) would be suitably modified, leading to a
faster-moving free boundary.

Note also that if the diffusion can be neglected,
(\ref{twosatsPDE}) and (\ref{twosatsbc1}) lead to $\frac 14
\Stil^{9/2} = \Qhat_{in} + \zhat - 1$ so (\ref{twosatsfbc}) gives
\begin{equation}
- \od\What\ttil = \frac {16}3 \left( \Qhat_{in} + \What - 1
\right) \, . \label{twosatsfbcm}
\end{equation}

The free-boundary condition (\ref{twosatsfbc}) only applies for an
advancing wet region, \\
$\dd\What/\dd\ttil \le 0$. An alternative
form is needed for when this region contracts, which will happen
when the rainfall decreases sufficiently. In any part of the soil
between the lowest location of the free boundary and its current
position, the roots can continue to remove water from the pellets,
thereby reducing $S_P$.\\

\medskip
As described in this paper we could now have at least four
types of region within the soil layer:
\begin{enumerate}
\item Dry zone, where $S = S_P = 0$; \item Damp or moist
(unsaturated) zone I, where $S= 0$, $0 < S_P < \phi_P$; \item Damp or
moist (unsaturated) zone II, where $0 < S < 1$, $S_P = \phi_P$; \item
Wet (saturated) zone, where $S = 1$, $S_P = \phi_P$.
\end{enumerate}


\section{Conclusions}
In this paper a one-dimensional time-dependent mathematical model
has been described for the development of the saturation in the
soil layer of a flat green roof. Our model suggests that a
fully saturated ($S=1$) region forms at the base of the soil layer
and this region can be thin relative to the total soil thickness.

From an initial dry state and from the onset of persistent rain,
fronts of saturation were computed to descend through the layer.
The decrease of saturation from unity following a decrease in
rainfall was also described. The end result is that most of the
rainwater falls through the soil layer and exits through the
network of holes in the bottom supporting sheet.

On a smaller scale, the pellets and soil particles are themselves
porous and made up of micropores. The water flow in and out of a
typical particle is modelled using the flux between (a) the
macropores (whose saturation is as modelled above) and (b) the
root system. This two-porosity model suggests that during the time
between spells of rain the micropores can retain (for long periods
of time) water that is available to be taken up by the roots.
For green roof design it is important to ensure
that the membrane supporting the soil is sufficiently permeable
to prevent any risk of full saturation. It is also important
to use soil which has sufficient micro-pores to soak up
large quantities of water during rainfall and
allow slow release during dry periods.

Further work might include adapting the soil thickness $L$ to
rainfall at the site of the building with the aim of making $L$ as
small as possible, while avoiding problems with saturation and
aridity. A first step towards this goal would be to carry out
experiments to more accurately determine the values of the
constants. Further simulations using more extensive rainfall data
could then be carried out to determine the optimum soil thickness.
In addition small modifications could be made to include the
influence of a sloped roof.\\

\medskip

\noindent\selectfont{\large{\textbf{Acknowledgements}}}\\
We acknowledge the support of the Mathematics Applications
Consortium for Science and Industry (www.macsi.ul.ie) funded by
the Science Foundation Ireland mathematics initiative grant
06/MI/005.

\end{document}